\documentclass[%
 reprint,
 amsmath,amssymb,
 aps,
]{revtex4-1}

\usepackage{graphicx}
\usepackage{dcolumn}
\usepackage{bm}

\begin{document}

\preprint{APS/123-QED}

\title{Quaternion Time, Mass, and Particle Velocity}
\thanks{Thanks to:  Prof. Leon Altschul, Prof. Shlomo Ruschin, Prof. Charles Schwartz, Dr. Daniel Alayon-Solarz, and Heidi Brun for
	helpful discussions}%
\author{Viktor Ariel}
\affiliation{%
}%

\begin{abstract}
In this work, we propose using real quaternions for the definition  of the time interval resulting in an alternative formulation of the relativistic space-time. We proceed with the quaternion definition of the particle mass that we derive using the mass-energy equivalence. This results in the quaternion mass dispersion relation, which is equivalent to the relativistic energy-momentum dispersion, however,  with a positive quadratic norm of the resulting quaternion space. Then, we define the quaternion particle velocity, which can be applied to relativistic particles and electrons in non-parabolic solid-state materials.  We show that the relativistic Lorentz expressions apply only to the absolute values of the time, mass and velocity while the physical 4-vector quantities are described by real quaternions. Finally, we introduce the quaternion electromagnetic potential and demonstrate that it has an effect on the particle mass and velocity. In particular, we show a direct connection between the electrostatic potential and the particle rest mass.
We compare the present model to experimental measurements of the electron velocity in InSb where we demonstrate quasi-relativistic electron velocity saturation at high electric fields. Therefore, it appears that physical quantities can be expressed by real quaternions while their measured observables and relativistic Lorentz transformations correspond to the quaternion absolute values. It seems that the quaternion approach presented here can be used as a basis for an alternative mathematical description of particle physics. \\
\end{abstract}

\pacs{Valid PACS appear here}
\maketitle


\section{\label{sec:level1}Introduction\protect\\}  
 
Hamilton, who invented quaternions \cite{Hamilton}, suggested that quaternion mathematics is the fundamental language of physics \cite{Gsponer}. Hamilton conceived that time is one-dimensional but space has three dimensions. Unfortunately, the Lorentz transformation was not discovered during Hamilton's lifetime, and eventually  Minkowski's 4-vector approach was used in relativistic mechanics. 

An apparent advantage of the quaternion formulation over the Minkowski approach is that real quaternions form a division algebra with a positive quadratic norm. This allows the use of the same quaternion framework in the description of both translational and rotational motion of particles, eliminating the need for vectors, tensors, and complex numbers. 

Previously, bi-quaternions were used in the description of relativistic mechanics \cite{Gsponer} and \cite{Silberstein} applying Minkowski's 4-vector norm. Unlike real quaternions however, bi-quaternions do not form a division algebra and consequently we avoid using them in the present work. 

First, we introduce the concept of quaternion time and show that the relativistic space-time can be conveniently represented by the resulting quaternion formulation with a positive quadratic norm, unlike in Minkowski's space-time \cite{Schwartz}.

It was demonstrated both experimentally and theoretically that electrons in non-parabolic isotropic semiconductors can be described by the quasi-relativistic energy-momentum dispersion relation \cite{Zawadzki}.
However, it is challenging to define the particle effective mass in solid state anisotropic materials where the vector velocity and momentum are not necessarily collinear.
We can not use a simple vector definition, $\vec{m} = \vec{p} / \vec{v}$, due to the lack of a well defined vector division in three-dimensional space. While it is possible to introduce a tensor effective mass, this seems to contradict experimental results showing that the measured mass is always a scalar quantity \cite{ArielEnergyBand}, \cite{ArielAnisotropic}.  

Previously, real quaternions were used to describe the free relativistic particle mass \cite{Gupta}. Similarly, we derive the particle mass using real quaternions from the relativistic dispersion relation and mass-energy equivalence  \cite{Jammer},  \cite{Okun}.  

We define the quaternion particle velocity as a function of both the quaternion time and quaternion mass. We show that absolute values of the quaternion time, mass, and velocity correspond to the correct relativistic Lorentz expressions.

Finally, we introduce the quaternion electromagnetic potential and observe its effect on the quaternion rest mass and particle velocity. We present a comparison between experimental measurements and the present model for the electron velocity in InSb at high electric fields  \cite{Asaukas}.

\section{Quaternion Time Interval} %

In his work, a real quaternion, $\textit{\textbf a}$, and its conjugate, $\bar{\textit{\textbf a}}$, are written in terms of a scalar, $a_0$, and an imaginary vector, $\vec{a}$,

\begin{equation}
\begin{cases}
{\textit{\textbf a}}  =a_0+ \thinspace\vec{a} = 
a_0+ \vec{i_1}\thinspace a_1 +  \vec{i_2}\thinspace a_2 + \vec{i_3}\thinspace a_3 \thinspace 
\\
\bar{{\textit{\textbf a}}}  =a_0- \thinspace\vec{a} = 
a_0- \vec{i_1}\thinspace a_1 -  \vec{i_2}\thinspace a_2 - \vec{i_3}\thinspace a_3 \thinspace 
\end{cases}
\label{eq:define_quaternion}
\end{equation}
The Euclidian imaginary vector basis is defined according to Hamilton \cite{Hamilton} as $\vec{i}\thinspace_1^2=\vec{i}\thinspace_2^2=\vec{i}\thinspace_3^2= \vec{i}_1\thinspace \vec{i}_2\thinspace \vec{i}_3 = -1$. The Euclidean norm of a quaternion is a scalar quantity defined in terms of the quaternion inner product as,

\begin{equation}
a = \|{\textit{\textbf a}}\|
=\sqrt{ \textit{\textbf a}\cdot \bar{\textit{\textbf a}}}\thinspace
=\sqrt{ \bar{\textit{\textbf a}}\cdot {\textit{\textbf a}}}\thinspace . 
\label{eq:define_abs}
\end{equation}

In general, we write quaternions as bold characters and their absolute values and scalars as regular characters. We use the vector sign above vectors using Gibbs-Heaviside notation, however, all vectors in this work are considered imaginary as originally intended by Hamilton.

Let us use the Lorentz transformation to express the absolute values of the space and time intervals for two moving frames,
\begin{equation}
c^2 t^2 - {x} ^2 = 
c^2 t^{\prime 2} -{x}^{\prime 2} \thinspace.
\label{eq:two_frames}
\end{equation}  
Here $t$ and $t^\prime$ are the absolute values of the time interval between two events and $x$ and $x^\prime$ are the absolute values of the space interval between the same events measured in two different moving frames. In relativistic mechanics, the absolute value of the saturation velocity is the speed of light in vacuum, $c$, which we consider a scalar constant thus limiting ourselves here to isotropic environments only.

In a special case, where one of the frames is at rest, the transformation (\ref{eq:two_frames}) can be written as,
\begin{equation}
c^2 t^2 - x^2 = c^2 t_0^{2}\thinspace,
\label{eq:rest_frame}
\end{equation}
where $t_0$ is the absolute value of the time interval in the rest frame. In (\ref{eq:rest_frame}), the absolute value of the time interval in the moving frame is $t$. The absolute value of the space interval is $x$, which is equal to the distance traveled by the moving frame relative to the rest frame during the time interval. 

Rearranging (\ref{eq:rest_frame}), we can write for the time interval in the moving frame,
\begin{equation}
t^2= t_0^2 + \dfrac {x^2}  {c^2}\thinspace.
\label{eq:moving_frame}
\end{equation}

Based on (\ref{eq:moving_frame}), we suggest that the time interval can be defined as a real quaternion,

\begin{equation}
{{\textit{\textbf t}}} =  t_0
+\thinspace \dfrac {\vec x } {c}  \thinspace .
\label{eq:define_quat_time_space}
\end{equation}
Therefore, the quaternion time interval consists of a scalar time interval measured in the rest frame and an imaginary space vector describing the distance traveled by the moving frame relative to the rest frame divided by the speed of light.

Note from (\ref{eq:define_quat_time_space}) that for slow moving frames, ${t_0 \gg x/c}$, we obtain a real scalar time, ${\textit{\textbf t}} \simeq t_0 $. However, for fast moving frames, $t_0 \ll x/c$, we obtain a pure quaternion time, ${\textit{\textbf t} } \simeq \thinspace  \vec{x} / {c}$, which is just an imaginary space vector measured in light-years. 

We could have used (\ref{eq:moving_frame}) to define the quaternion space rather than time. However, we prefer to use time as the more fundamental quantity because, in the rest frame, where $x=0$, both (\ref{eq:moving_frame}) and  (\ref{eq:define_quat_time_space})  describe the rest time. Therefore, the rest time interval, $t_0$, appears to be the most fundamental physical quantity and an invariant for all moving frames.

We calculate the norm of the quaternion time from (\ref{eq:define_abs}) and (\ref{eq:define_quat_time_space})
\begin{equation}
t^2
= {\textit{\textbf t}}\cdot \bar{\textit{\textbf  {t}}} 
= \bar{{\textit{\textbf t}}}\cdot {\textit{\textbf  {t}}} 
= t_0^2 + \dfrac { {x}\thinspace^2}  {c^2}\thinspace,
\label{eq:space_time_sq}
\end{equation}
which is the same as the Lorentz expression (\ref{eq:moving_frame}).
Note that the quadratic norm of the quaternion time in (\ref{eq:space_time_sq}) is a positive quantity. In Minkowski's 4-dimensional space-time, the quadratic norm is the same as the square of the rest time interval in our formulation, 
\begin{equation}
t_0^2= t^2 - \dfrac {x^2}  {c^2}\thinspace.
\label{eq:mink_norm}
\end{equation}
This expression can be potentially negative leading to purely theoretical particles moving faster than the speed of light and known as tachyons \cite{Schwartz}.

Alternatively, we can use the quaternion time in (\ref{eq:define_quat_time_space}) to define a location of a point in the relativistic space-time. Since real quaternions can describe both translational and rotational motions, the quaternion formulation can be applied to arbitrary moving frames and not only to inertial frames of reference.

Therefore, we have defined the quaternion time interval which can be used as an alternative formulation of the relativistic space-time.

\section{Quaternion Mass} %

Let us assume that a free relativistic particle, or a particle in non-parabolic semiconductors \cite{Zawadzki}, \cite{ArielEnergyBand}, \cite{ArielAnisotropic}
can be described by the energy-momentum dispersion relation,  
\begin{equation}
E^2 =  m_0^2 c^4 + {p^2} c^2 \thinspace.
\label{eq:energy_momentum}
\end{equation}
Here $E$ is the total energy, $m_0$ is the scalar rest mass, and ${p}\thinspace$ is the absolute value of the particle momentum due to particle movement relative to the rest frame.

Next, we apply Einstein's mass-energy equivalence, ${E=m c^2}$. Note that Einstein often used an alternative formulation ${E_0=m_0 c^2}$, where $E_0$ is the energy of the body at rest. This led to a debate about the physical meaning of the mass-energy equivalence \cite{Jammer}, \cite{Okun}. Here, we assume the validity of both expressions as necessary to explain experimental measurements of particles at rest and particles with a 
zero rest mass, like electrons in graphene and photons \cite{ArielEnergyBand}.  Therefore using ${E=m c^2}$, we obtain from  (\ref{eq:energy_momentum}),

\begin{equation}
m^2 =  m_0^2+ \dfrac {{p}\thinspace ^2} {c^2} \thinspace .
\label{eq:mass-momentum}
\end{equation}

Based on (\ref{eq:mass-momentum}), we propose the following definition of the quaternion mass, which we assume is a function of the quaternion time interval,

\begin{equation}
{\textit{\textbf m}}({\textit{\textbf t}}) 
= m_0({\textit{\textbf t}})+ \thinspace \dfrac {\vec p ({\textit{\textbf t}})} {c} 
\thinspace .
\label{eq:q_mass_define}
\end{equation}

\begin{figure}
	\includegraphics{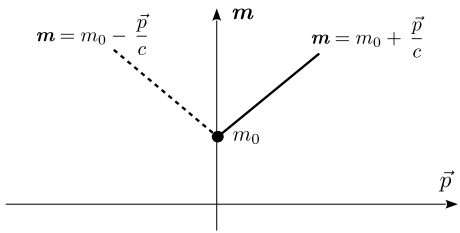}
	\caption{\label{fig:mass-momentum} A two-dimensional representation of the quaternion mass and its conjugate as functions of vector momentum.}
\end{figure}
\begin{figure}
	\includegraphics{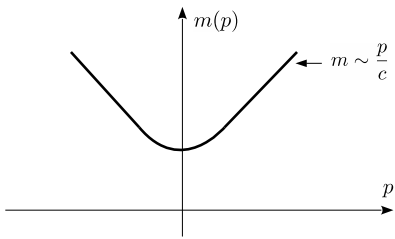}
	\caption{\label{fig:abs-mass-momentum} The absolute value of the quaternion mass as a function of the absolute momentum.}
\end{figure}
Note from (\ref{eq:q_mass_define}) that for slow moving particles, ${m_0 \gg p/c}$, we obtain a real scalar mass, ${\textit{\textbf m}} \simeq m_0$. For fast moving particles like photons  and electrons in graphene, $m_0 \ll p/c$, we obtain the pure imaginary vector mass, ${{\textit{\textbf m}} \simeq  \thinspace {\vec{p}} / {c}}$. 

Also, note that we could have used (\ref{eq:energy_momentum}) to define the quaternion energy or quaternion momentum. However, we prefer to focus on the particle mass here, because for $p=0$, expressions (\ref{eq:energy_momentum}), (\ref{eq:mass-momentum}), and (\ref{eq:q_mass_define}) describe the particle at rest, where the rest mass can be easily measured. 

Using (\ref{eq:q_mass_define}), we can calculate the norm of the quaternion mass,
\begin{equation}
m^2=
{\textit{\textbf m}}\thinspace\bar{{\textit{\textbf m}}}= 
\bar{{\textit{\textbf m}}}\thinspace{{\textit{\textbf m}}}= 
m_0^2+ \dfrac {{p}\thinspace ^2} {c^2} \thinspace.
\label{eq:mass_dispersion}
\end{equation}
Therefore, we transformed the relativistic energy-momentum dispersion relation (\ref{eq:energy_momentum}) into the quaternion mass dispersion relation (\ref{eq:mass_dispersion}) using Einstein's mass-energy equivalence.

We identify the square of the mass absolute value in (\ref{eq:mass_dispersion}) as the positive quadratic norm of the quaternion mass configuration space.

In Fig.~\ref{fig:mass-momentum}, we demonstrate a linear dependence of the quaternion mass on  particle momentum from (\ref{eq:q_mass_define}). For comparison in Fig.~\ref{fig:abs-mass-momentum}, we show the absolute value of the quaternion mass from  (\ref{eq:mass_dispersion}) and notice the familiar behavior similar to the energy bands in non-parabolic semiconductors  \cite{ArielEnergyBand}. We suggest that the quaternion mass in Fig.~\ref{fig:mass-momentum} represents the correct mathematical and physical descriptions of the particle mass while its absolute value in Fig.~\ref{fig:abs-mass-momentum} corresponds to experimentally observable results.

Therefore, we have defined the quaternion mass composed of a scalar particle rest mass and an imaginary vector mass due to the particle motion relative to the rest frame. We showed that the absolute value of the quaternion mass leads to the mass dispersion relation and the quaternion mass configuration space with a positive quadratic norm.

\section{Quaternion Velocity} %
 
In classical mechanics, the particle velocity is usually defined in terms of the vector space interval and the scalar time interval. Using the quaternion time (\ref{eq:define_quat_time_space}), we can define the quaternion velocity by analogy with classical mechanics,

\begin{equation}
{\textit{\textbf v}}({\textit{\textbf t}})= \dfrac {\thinspace \vec{x}} {\textit{\textbf {t}}} \thinspace .
\label{eq:velocity_define}
\end{equation}
This definition is mathematically possible only because division by the quaternion time is well defined for real quaternions.

Quaternion division is not commutative, which leads to two possible solutions for the particle velocity (as shown in the Appendix),

\begin{equation}
{\textit{\textbf {v}}} =
\begin {cases} 
\dfrac  {\vec{x} \thinspace \bar{\textit{\textbf{t}}}} {{t}^2}
= \dfrac  {1}  {t^2} 
\left(\dfrac {\vec{x} \cdot \vec {x}} {c}\thinspace 
+ \thinspace t_0\vec{x} -\dfrac {\vec{x}\times \vec{x}} {c}\thinspace\right)
\\
\\
\dfrac  {\bar{{\textit{\textbf t}}}\thinspace {\vec{x}} } {{t}^2}
= \dfrac  {1}  {t^2} 
\left(\dfrac {\vec{x} \cdot \vec {x}} {c}\thinspace 
+ \thinspace t_0\vec{x} +\dfrac {\vec{x}\times \vec{x}} {c}\thinspace\right)
\end {cases}
\label{eq:q_t_velocity}
\end{equation}

However, the vector product in (\ref{eq:q_t_velocity}) vanishes because $\vec{x} \times \vec{x} =0$. Therefore, the multiplication in (\ref{eq:q_t_velocity}) is commutative and we obtain the following single expression for the quaternion particle velocity,

\begin{equation}
{\textit{\textbf v} }
= \dfrac {1} {t^2 } \left( \dfrac {{x\thinspace}^2} {c}
+\thinspace  {t_0  \thinspace\vec{x} }   \thinspace \right )
= \dfrac {x^2} {c t^2} +\thinspace  
\dfrac { t_0 \vec{x} } {t^2}
\thinspace .
\label{eq:q_velocity_final}
\end{equation}
As can be seen from (\ref{eq:q_velocity_final}), the quaternion velocity has a non-zero scalar part, which vanishes only for particles at rest,  ${\textit{\textbf v}} = 0$ when ${x} =0 $.

Also by analogy with classical mechanics, we can define the quaternion velocity using the quaternion particle mass (\ref{eq:q_mass_define}).  Using the quaternion division approach similar to (\ref{eq:q_t_velocity}), we obtain:

\begin{equation}
{\textit{\textbf v}} = \dfrac {\vec{p}} {\textit{\textbf m}} 
= \dfrac {p^2} {m^2 c} +\thinspace  
\dfrac { m_0 \vec{p} } {m^2}
\thinspace .
\label{eq:velocity_define_mass}
\end{equation}

We assume that both definitions of the quaternion velocity, (\ref{eq:velocity_define})  and  (\ref{eq:velocity_define_mass}), produce the same result,

\begin{equation}
{\textit{\textbf v}} =\dfrac {\vec {p}} {\textit{\textbf m}} = \dfrac {\vec{x}} {\textit{\textbf t}}\thinspace .
\label{eq:velocity_equivalence}
\end{equation}

Also, we can write for the absolute values from (\ref{eq:velocity_equivalence}),

\begin{equation}
{v} =\dfrac {{p}} { {m}} = \dfrac {{x}} {{t}}\thinspace .
\label{eq:velocity_abs_equivalence}
\end{equation}

Therefore, we obtain from (\ref{eq:velocity_equivalence}) and (\ref{eq:velocity_abs_equivalence}),
\begin{equation}
{\textit{\textbf v}} = \dfrac {v^2} { c}+\dfrac { t_0 \vec{x} } {t^2} 
= \dfrac {v^2} { c}+\dfrac { m_0 \vec{p} } {m^2}\thinspace .
\label{eq:velocity_new_equivalence}
\end{equation}
The above implies that vector momentum $\vec{p}$ is collinear with the space interval $\vec{x}$. However the quaternion velocity has a scalar part which only vanishes when $v=0$.

Calculating for the absolute value of the quaternion time from (\ref{eq:velocity_new_equivalence}) and using $x = vt$ from (\ref{eq:velocity_abs_equivalence}), 
\begin{equation}
{t} = \dfrac {t_0} {\sqrt {1-\dfrac {v^2} {c^2}}} \thinspace ,
\label{eq:rel_abs_time}
\end{equation}
which we recognize as the Lorentz time transformation. 

Similarly, for the absolute value of the quaternion mass from (\ref{eq:velocity_new_equivalence}) and using $p=m v$ from (\ref{eq:velocity_abs_equivalence}), 
\begin{equation}
{m} = \dfrac {m_0} {\sqrt {1-\dfrac {v^2} {c^2}}} \thinspace ,
\label{eq:rel_abs_mass}
\end{equation}
which is the Lorentz transformation of mass.

Rearranging  (\ref{eq:rel_abs_time}) and (\ref{eq:rel_abs_mass}) we obtain the Lorentz expressions for the absolute value of the velocity,
\begin{equation}
{v} = {c} \thinspace {\sqrt {1-\dfrac {t_0^2} {t^2}}}
= {c} \thinspace {\sqrt {1-\dfrac {m_0^2} {m^2}}} 
\thinspace .
\label{eq:rel_abs_velocity}
\end{equation}
Expressions (\ref{eq:rel_abs_velocity}) clearly show velocity saturation for particles with a relatively small rest mass, $m_0 \ll m$, and a zero velocity for particles at rest, when $t=t_0$ and $m=m_0$.

Therefore, we were able to define the quaternion velocity by using both the quaternion time and mass and showed that the absolute values of the quaternion time, mass and velocity correspond to their traditional Lorentz expressions.

\section{Mass and Velocity in Electromagnetic Potential} %

Next, we introduce an electromagnetic interaction and calculate its influence on the quaternion mass and velocity.  We define the quaternion electromagnetic potential using the relativistic 4-vector analogy,

\begin{equation}
{\boldsymbol \phi}({\textit{\textbf t}}) 
=\phi+\thinspace\thinspace c {\vec A}  \thinspace.
\label{eq:em_field}
\end{equation}
Here $\phi$ is scalar electrostatic potential and $\vec A$ is the electromagnetic vector potential defined similar to the traditional relativistic approach.

We propose the following definition  of the quaternion mass in the presence of an electromagnetic interaction,
\begin{equation}
{\textit{\textbf m}}_{e} ({\textit{\textbf t}},{\boldsymbol \phi} ) 
 = {\textit{\textbf m}}({\textit{\textbf t}})
 -\dfrac {q \boldsymbol \phi({\textit{\textbf t}})}  {c^2} \thinspace ,
\label{eq:em_mass}
\end{equation}
where $q$ is the electron charge.
This leads to the following expression for the total quaternion mass in the presence of an electromagnetic interaction, using (\ref{eq:q_mass_define})  and (\ref{eq:em_mass}),
\begin{equation}
{\textit{\textbf m}}_{e}   
= m_0-\dfrac {q{\phi}} {c^2} 
+\dfrac {\vec p-q\vec A} {c} \thinspace .
\label{eq:total_mass}
\end{equation}

It appears that electrostatic scalar potential, $\phi$, has an effect on the rest mass while the electromagnetic vector potential, $\vec A$, affects the vector momentum. 

We can now calculate the square of the mass absolute value under the influence of the electromagnetic potential as,

\begin{equation}
m_e^2
=\left( m_0-\dfrac {q{\phi}} {c^2}\right)^2 + \dfrac { |\vec p-q\vec A |^2} {c^2} 
\thinspace .
\label{eq:total_mass_square}
\end{equation}
Assuming the validity of ${m_e}\thinspace {v} =|\vec p-{q}  \vec A | $,  we can calculate the absolute value of quaternion mass in the presence of an electromagnetic interaction,

\begin{equation}
{m_e} = \dfrac {{m_0}-\dfrac {q {\phi}} {c^2}} {\sqrt {1-\dfrac {v^2} {c^2}}}\thinspace .
\label{eq:total_abs_mass}
\end{equation}
Comparing (\ref{eq:total_abs_mass}) to (\ref{eq:rel_abs_mass}), we clearly see the impact of electrostatic potential on particle mass.

Including an electromagnetic interaction into (\ref{eq:velocity_define_mass}), we obtain the following quaternion velocity expression,
\begin{equation}
{\textit{\textbf v}}_e 
=\dfrac   {{|\vec p-{q}  \vec A |}^2} {m^2 c}
+ \left ({m_0} 
- \dfrac {q {\phi}} {c^2} \right ) \dfrac {\vec p-{q} \vec A} {m^2 }.
\label{eq:em_velocity}
\end{equation}

\begin{figure}
	\includegraphics{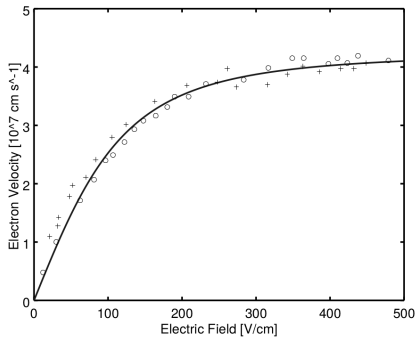}
	\caption{\label{fig:sat_velocity} Saturation velocity in InSb obtained from measurements (symbols) \cite{Asaukas} and compared to the present model (solid line).}
\end{figure}

Finally, we can calculate the absolute value of the velocity from (\ref{eq:em_velocity}),
\begin{equation}
v_e=
\dfrac {c } 
{\sqrt {1+\left( m_0 c-\dfrac {q \phi} {c}\right)^2  /  \thinspace
		|\vec p-{q}  \vec A |^2}}\thinspace.
\label{eq:em_abs_velocity}
\end{equation}
Comparing (\ref{eq:em_abs_velocity}) to (\ref{eq:rel_abs_velocity}), we can clearly see the influence of both the electrostatic potential and vector electromagnetic field on the particle velocity. 

Note that in our approach, both the electrostatic potential, $\phi$, and the rest mass, $m_0$, are generally distributed functions of the quaternion time, {\textit{\textbf t}}. Therefore using the point particle approximation, we may consider the scalar rest mass as an average effective mass located at the center-of-mass position and representing the influence of the internal structure and external interactions, and thus allowing description of any physical object from (\ref{eq:total_mass}). Similarly, the average effect of external interactions on the particle velocity can be considered as a modification of the effective particle saturation velocity in (\ref{eq:em_abs_velocity}).

For example, we may consider electrons and holes in solid-state materials as regular relativistic particles, using however the modified rest mass and saturation velocity due to the influence of the crystal lattice electromagnetic potential \cite{ArielEnergyBand}. Such quasi-relativistic behavior was predicted theoretically and demonstrated experimentally in InSb \cite{Zawadzki}. Since the saturation velocity in typical solid-state materials is several orders of magnitude smaller than the speed of light, we can expect a profound effect of the external electromagnetic potential on electron velocity in non-parabolic semiconductors.

In Fig.~\ref{fig:sat_velocity}, we present a comparison between the present model (\ref{eq:em_abs_velocity}) and experimental measurements of the electron velocity in InSb  \cite{Asaukas}, using $m_0$ as the electron effective mass and $c$ as the saturation velocity. We used the electron life-time as a fitting parameter in order to calculate the momentum from the applied electric field. As can be seen in Fig.~\ref{fig:sat_velocity}, there is close correspondence between the experimental measurements and our quasi-relativistic velocity model.

Therefore, we introduced the quaternion electromagnetic potential and demonstrated its effect on the particle mass and velocity. In particular, the electrostatic potential has a direct effect on the particle rest mass. We showed that quasi-relativistic velocity saturation can be demonstrated in non-parabolic materials such as InSb.

\section{Conclusions}

In this work, we introduced the real quaternion time interval, which consists of a scalar rest-time plus an imaginary space vector, as an alternative formulation of the relativistic space-time.

Then, we used the relativistic energy-momentum dispersion relation and the mass-energy equivalence to define the quaternion mass. We showed that the quaternion mass consists of a scalar rest mass and an imaginary vector momentum due to particle motion. We showed that the energy-momentum dispersion relation is transformed into the quaternion mass dispersion relation with a positive quadratic norm. Since the quaternion mass is a function of the quaternion time, we assume that the quaternion mass can be used for the description of matter in relativistic space-time.

We defined the quaternion particle velocity and demonstrated that it can be uniquely expressed as a function of both the quaternion time and quaternion mass. Unlike the traditional vector velocity, the quaternion velocity has a non-vanishing scalar component.
We showed that the absolute values of the quaternion time, mass, and velocity correspond to their relativistic Lorentz equivalents.

Most importantly, we used the quaternion electromagnetic potential to derive the quaternion mass and velocity in the presence of an electromagnetic interaction. It appears that the electromagnetic potential has a direct effect on the rest mass and particle velocity. We compared the theoretical and experimental velocity absolute value in InSb and showed that our model can accurately predict the quasi-relativistic velocity saturation.

It appears that quaternions can be used for the description of physical quantities while their absolute values correspond to the experimental observables and Lorentz transformations. 
Finally, the real quaternions form a division algebra with a positive quadratic norm and therefore seem suitable for the unified mathematical description of time and matter, justifying Hamilton's conjecture that quaternions are the fundamental language of physics.

\section{Appendix}

Quaternion division is not commutative and is defined for two quaternions ${\textit{\textbf a}}$ and ${\textit{\textbf b}}$ as,
\begin{equation}
\begin {cases} 
{\textit{\textbf a}}\thinspace {\textit{\textbf b}}^{-1} =\dfrac  {{\textit{\textbf a}} \thinspace\bar{{\textit{\textbf b}}}} {{b}^2}
=\dfrac  {1}  {b^2} 
\left[b_0 a_0 +\vec{b} \cdot \vec {a}\thinspace 
+ \thinspace(b_0\vec{a} - a_0 \vec {b} 
-\vec{a}\times \vec{b}\thinspace)\right]
\\
\\
{\textit{\textbf b}}^{-1}\thinspace  {\textit{\textbf a}} =\dfrac  {\bar{{\textit{\textbf b}}} \thinspace{\textit{\textbf a}}} {{b}^2}
=\dfrac  {1}  {b^2} 
\left[b_0 a_0 +\vec{b} \cdot \vec {a}\thinspace 
+ \thinspace(b_0\vec{a} - a_0 \vec {b} 
+\vec{a}\times \vec{b}\thinspace)\right]
\end {cases}
\label{eq:quaternion_div}
\end{equation}
\pagebreak

\end{document}